\documentclass[5p,times]{elsarticle}

\usepackage[T1]{fontenc}
\usepackage{graphicx}
\usepackage{dcolumn}
\usepackage{bm}
\usepackage{url}
\usepackage[dvipsnames]{xcolor}
\usepackage{amsmath}
\usepackage{amssymb}
\usepackage{xpatch}

\usepackage[english]{babel}
\usepackage[colorinlistoftodos, color=green!40, prependcaption]{todonotes}
\usepackage{comment}
\usepackage[pdftex, pdftitle={Article}, pdfauthor={Author}]{hyperref} 

\def\cevns{{CE$\nu$NS}}
\xpatchcmd{\MaketitleBox}{\hrule\vskip12pt}{\vspace{-1\baselineskip}}{}{} 

\begin{document}

\begin{frontmatter}

\title{A portable neutron scatter camera with isotropic sensitivity}

\author[uchicago]{S.G.\ Yoon\fnref{equal}}

\author[dipc,EHU]{L. Larizgoitia\fnref{equal,now}}

\author[uchicago,dipc]{C.M.\ Lewis\corref{cor1}}
\ead{mark.lewis@dipc.org}

\author[dipc,ikerbasque]{F. Monrabal}

\author[uchicago,dipc,ikerbasque]{J.I.\ Collar}

\address[uchicago]{Enrico Fermi Institute, Kavli Institute for Cosmological
            Physics, and Department of Physics, University of Chicago, Chicago,
            Illinois 60637, USA}

\address[dipc]{Donostia International Physics Center, BERC Basque Excellence Research Centre, Manuel
            Lardizabal 4, 20018, Donostia-San Sebasti\'an, Spain}

\address[ikerbasque]{Ikerbasque, Basque Foundation for Science, Plaza
            Euskadi 5, 48013, Bilbao, Spain}

\address[EHU]{Department of Physics, University of the Basque Country UPV/EHU, PO Box 644, Bilbao, E-48080, Spain}

\cortext[cor1]{Corresponding author}
\fntext[equal]{These authors contributed equally to this work.}
\fntext[now]{Now at IDOM Consulting, Engineering, Architecture, S.A.U.}

\begin{abstract}

We describe the design and characterization of a  neutron scatter camera specifically developed for the study of neutron fields (flux, energy, direction) in experimental areas of spallation facilities, where multiple neutron sources can simultaneously contribute. A combination of hardware and logic triggers provides the isotropic response required for this application. Neutron source locations are accurately reconstructed on  images from a built-in  360$^\circ$ panoramic  camera. The full understanding of neutron fields that the device provides can be used for site selection and  abatement of neutron backgrounds affecting sensitive instruments such as  coherent elastic neutrino-nucleus scattering (\cevns) detectors to be located at the upcoming European Spallation Source. 
\end{abstract}

\end{frontmatter}

\section{Introduction}

The neutron scatter camera (NSC) is a  recently developed  radiation detector   capable of identifying the flux and energy  of local neutron fields in addition to providing imaging capabilities that allow for accurate reconstruction of neutron incoming direction and/or source location \cite{nc1,nc2,nc3,nc4,nc5,nc6,nc7}.  As such, these sophisticated devices have found a niche of applications in treaty verification, warhead monitoring, homeland security, and in general wherever the detection of special nuclear material is involved \cite{nc1,nc2}. Their principle of operation is similar to that of Compton gamma-ray cameras \cite{compton} for which multi-detector arrays provide directional information via conservation of energy and momentum. For a single incident particle this information is limited, but following a sufficient exposure the spatial direction of neutron origin(s) can be revealed. 

Spallation facilities provide the most intense neutron sources available for fundamental and applied research \cite{spallation}. Additionally, they generate pulsed neutrino fluxes of much interest for particle physics and in particular for the study of  coherent elastic neutrino-nucleus scattering (\cevns) \cite{sp1,sp2,sp3}. While significant measures against neutron leakage from the spallation target and adjacent instrument beamlines are taken in the interest of radioprotection, an extreme initial neutron hardness (hundreds of MeV) results in residual pulsed neutron fluxes sufficient to induce  backgrounds in \cevns\  detectors or  other neutron-sensitive instruments operated at the facility. 

A full simulation of neutron transport in the vicinity of spallation sources is a difficult endeavor \cite{lewis,leire}. Secondary sources neglected during this exercise can be dominant once on site. For instance, during the characterization of neutron backgrounds that preceded the first observation of \cevns\ at the SNS \cite{science,bjorn} it was noticed that sites along a basement "neutrino alley"  naively expected to be  increasingly  protected by shielding along the line of sight to  target, were in reality subject to a fast neutron flux growing by  orders of magnitude. The origin of these neutrons was traced to leakage along the proton beamline one floor above, with a point of ingress through a distant staircase. During our more recent work in the third-floor terrace of J-PARC's MLF \cite{sp1}, performed within the context of MLF experiment 2025A0375, we have characterized three spatially and temporally distinct neutron sources reaching a suitable \cevns\ experimental site. Similarly, triangular utility rooms at the upcoming ESS \cite{sp2,sp3} are an emplacement of interest for neutrino activities. A  room corner nearest  to the target monolith is suspected of neutron leakage \cite{lewis,leire}. If confirmed via  imaging and characterized in energy, the addition of remedial shielding could be contemplated. Nevertheless, existing staircases leading to floors above and below might generate a situation similar to that described at the SNS. 

Informed by our experience, we have revisited the NSC design paradigm, optimizing the technique for  situations where a multiplicity of neutron sources might be simultaneously present, while considering the specifics of  expected intensity and  hardness for stray neutrons  at spallation facilities. With the goal of free installation at a variety of ESS sites, we have kept in mind safety considerations by avoiding the use of  xylene-based liquid scintillator ---the default detector choice in NSCs--- instead favoring recently developed EJ-276D, a plastic alternative with sufficient neutron-gamma discrimination \cite{2761,2762}. The resulting portable NSC (Fig.\ \ref{fig:camera}) requires a single nearby electrical outlet for operation and features an innovative trigger mechanism that provides the desired full-4$\pi$ isotropic response, as well as a useful overlay of reconstructed  sources onto   images from a built-in 360$^{\circ}$ panoramic camera. In what follows we describe the optimization, design and tests of this NSC, confirming its readiness for facility-wide operation coincident with first protons-on-target (POT) at the ESS.

\section{Detector}

\subsection{Principle} \label{how}

\begin{figure}[!htbp]
\centering
\includegraphics[width=1.\linewidth]{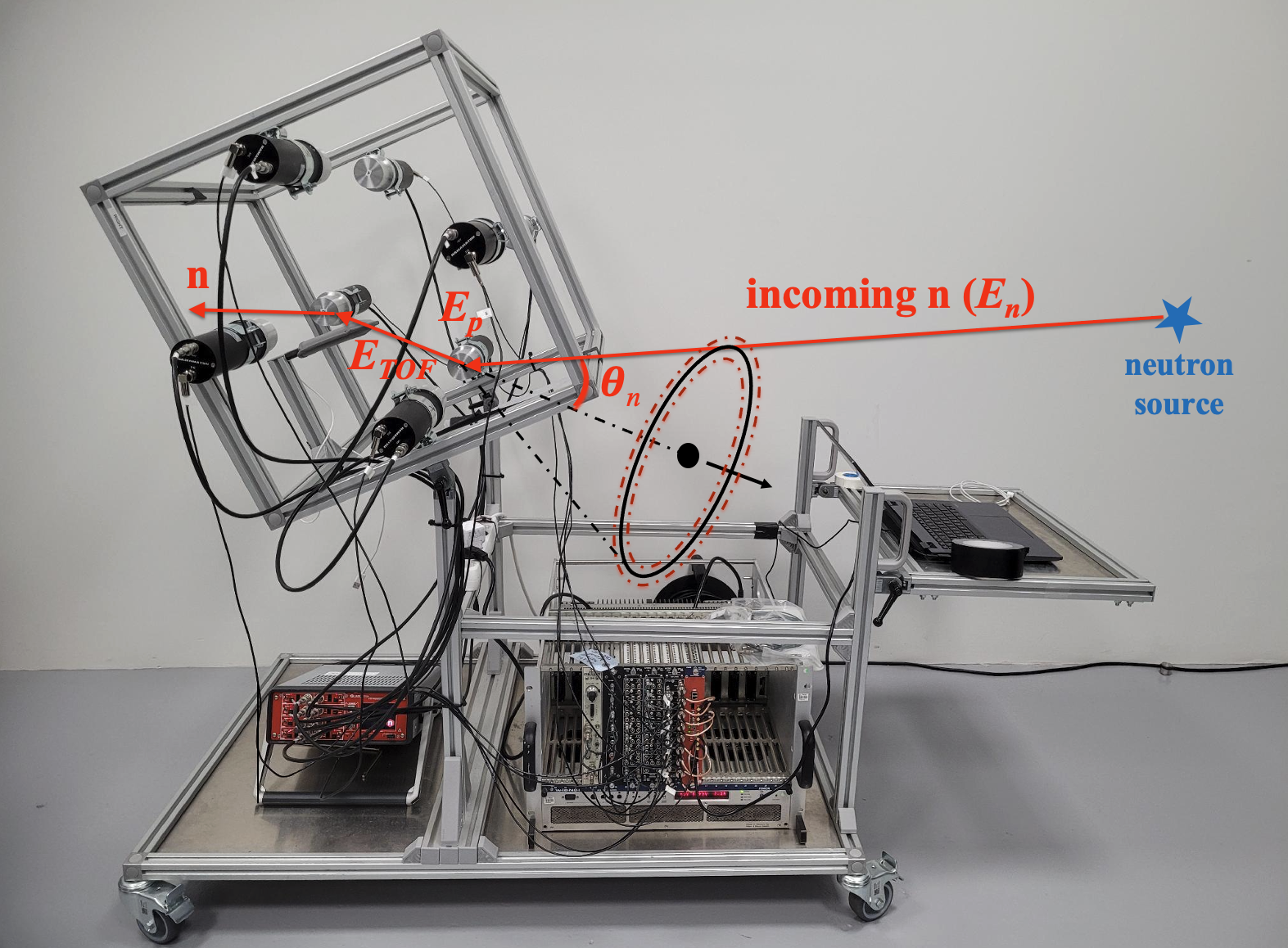}
\caption{\label{fig:camera} {Image of the portable NSC. Overlaid is an example of neutron interaction with two of eight detector cells. Labels denote incoming neutron energy ($E_n$), energy deposited in the first cell ($E_p$) and that carried by the neutron between interactions ($E_{TOF}$). Dashed  rings represent the uncertainty in the reconstructed scattering angle ($\theta_n$) due to finite position, time and energy resolutions. The emplacement of a  360$^\circ$ optical camera, visible near the second scattering point, does not obstruct rectilinear neutron trajectories between cells. }  }
\end{figure}

A majority of prior NSC designs rely on neutron coincidences involving two parallel planes of detectors \cite{nc1,nc3,nc5,nc6,nc7}. This generates a preferential forward-backward sensitivity in the device along an axis perpendicular to the planes. In high-flux situations involving a single neutron origin with suspected or known location this does not represent a limitation. Similarly, if time to positive identification is not of the essence, multiple runs with changing camera orientations can provide an enhanced spatial coverage. 

The specific needs of our application, described in the previous section, led us to explore a geometry involving eight EJ-276D cylindrical cells, read out by Hamamatsu H6410 photomultiplier (PMT) modules, forming a cube (Fig. \ref{fig:camera}). A small cell size (5 cm diameter by 5 cm length) was favored so as to reduce the possibility of multiple scattering within, as this unaccounted-for effect blurs angular and energy reconstruction. Nevertheless, this is accomplished at the expense of a reduced neutron detection efficiency. The simulations discussed below confirmed that for  expected neutron fluxes at the ESS this efficiency penalty is tolerable.  

The NSC concept exploits conservation of energy and momentum in elastic collisions between incident neutrons and protons in a hydrogenated detector \cite{nc5,nc6,nc7}. Incoming neutrons must scatter in one cell and  again in a second (Fig. \ref{fig:camera}). The energy deposited in the first ($E_p$) is measured directly and added to the energy carried by the  neutron between detectors ($E_{TOF}$), which is inferred by its time-of-flight (TOF),

\begin{equation} \label{eq:tof}
    E_{\text{TOF}} = \frac{1}{2}m_n \left ( \frac{d}{\text{TOF}} \right )^2
\end{equation}

\begin{equation} \label{eq:total_en}
    E_n = E_p + E_{\text{TOF}} \quad ,
\end{equation}
where $m_n$ is the mass of the neutron, $d$ the  separation between scatters and $E_n$ the reconstructed incoming neutron energy. As a reference, TOF (1 m) $=72.3 $ ns$/\sqrt{E_{TOF}\mathrm{(MeV)}}$. Since the direction of the recoiling proton is not measured, the trajectory of the incoming neutron can only be constrained to a cone around an axis joining the two scatters. The cone is defined by the initial scatter angle $\theta_n$ (Fig. \ref{fig:camera}), where 
\begin{equation} \label{eq:angle_recon}
    \tan^2{\theta_n} = \frac{E_p}{E_{\text{TOF}}} \quad .
\end{equation}
The finite size of scintillator volumes presents a range of possible values of $d$ for interactions involving a pair of cells. For each neutron interaction involving two cells the polar phase space subtended by the conical projection of $\theta_n$ is smeared by this effect. A finite resolution in the determination of $E_p$ and $E_{TOF}$ provides a second source of uncertainty in $\theta_n$ (Fig. \ref{fig:camera}). Following a sufficiently long exposure, a heatmap of summed back-projections of reconstructed cones reveals neutron source positions in angular coordinates using a suitable origin (Sec.\ \ref{position}).

\subsection{Design} \label{design}

\begin{figure}[!htbp]
\centering
\includegraphics[width=.95\linewidth]{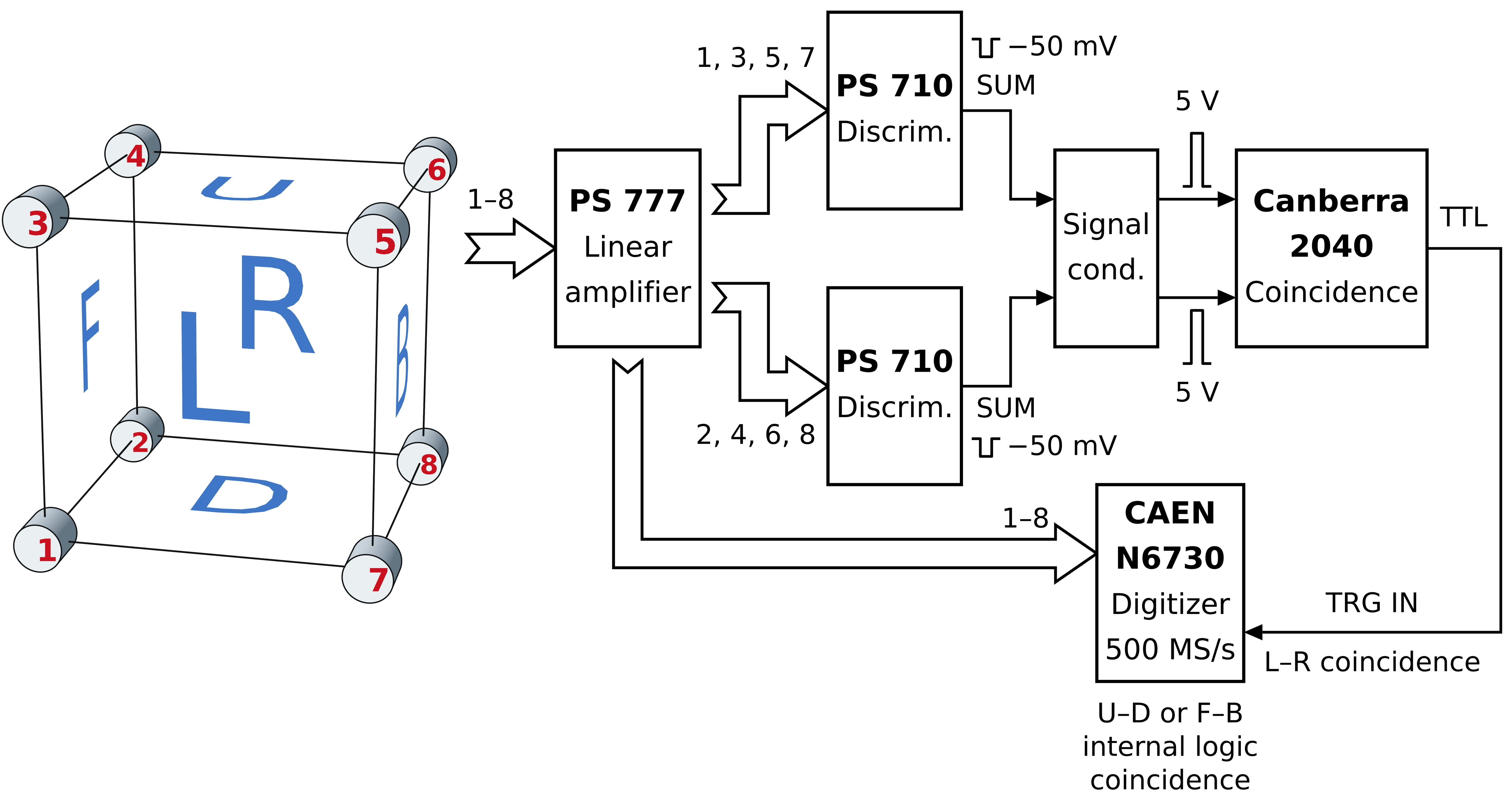}
\caption{\label{fig:daq} { Combination of internal logic and hardware triggers leading to the isotropic sensitivity of the NSC (see text). Phillips Scientific (PS), Canberra, and CAEN NIM modules are employed by this DAQ.   }  }
\end{figure}

Fig. \ref{fig:daq} depicts the spatial arrangement of detector cells and  three pairs of planes (Left-Right, Up-Down, Front-Back) involved in the generation of data-acquisition (DAQ) triggers. A {\it trigger condition }forces a CAEN N6730SB digitizer  to fetch waveforms from all its eight channels (0-7), one per cell, which are then bundled and saved to disk for analysis. The self-trigger logic of the N6730SB produces a {\it trigger request} whenever the digitized input from any channel surpasses a user-defined DC level,  set here to a $\sim$20 keV equivalent (see Sec. \ref{energy_cal} for a discussion on energy calibration). N6730SB-specific programmable settings (majority level, T$_{TVAW}$) can be used to restrict the appearance of a trigger condition to coincidences of at least two unique trigger requests within a time window. This is set to its maximum (120 ns) in our case, allowing for the detection of $E_{TOF}\gtrsim$ 60 keV. A repeated trigger request of the same type does not set off a trigger condition.   

Regrettably, the N6730SB provides only four unique trigger requests, each arising from the self-trigger of either detector cell feeding a pair of consecutive digitizer channels: TRG\_REQ[1] for channels 0-1, corresponding to cell 1 or cell 2 firing, TRG\_REQ[2] for channels 2-3 whenever cell 3 or cell 4 self-triggers, etc. As can be noticed from Fig. \ref{fig:daq}, this limitation (four unique trigger requests for eight channels) would impair  lateral (L-R) NSC sensitivity, especially for fast neutrons prone to forward scattering. For instance, a neutron originating on the left side of the NSC and interacting with cells 5 and 6, corresponding to adjacent N6730SB channels 4 and 5, both producing the same TRG\_REQ[3], would not reach the minimum of two different, unique trigger requests necessary for a coincidence and the ensuing trigger condition.

This deficiency is compensated for by additionally enabling trigger conditions from an external hardware trigger (TRG IN) derived from coincidences involving the L and R planes (Fig. \ref{fig:daq}). A signal conditioning NIM module inverts and amplifies sum (i.e., OR) outputs from PS 710 discriminators, one each for L and R cell planes. The circuit is based on the LM7171 Texas Instruments operational amplifier, able to provide a very high slew-rate (4100 V/$\mu$s). This property is needed to satisfy the $<$50 ns rise-time, >4 V specification at the inputs of the Canberra 2040 coincidence unit employed. This combination of hardware and internal logic trigger conditions provides the isotropic response sought from this neutron scatter camera. 

\begin{figure}[!htbp]
    \centering
    \includegraphics[width=.9\linewidth]{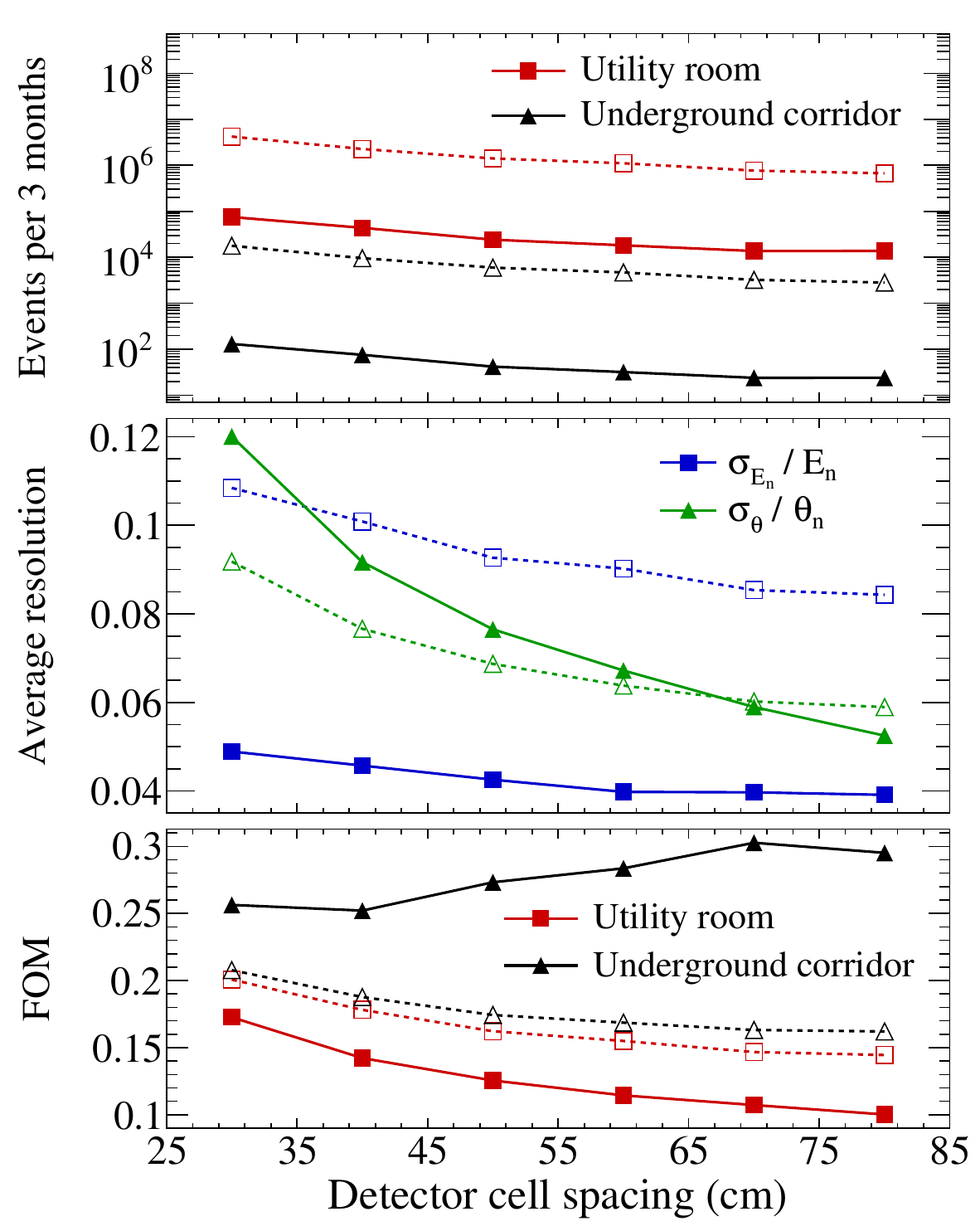}
    \caption{
        \label{fig:design}
        MCNPX-PoliMi optimization of  detector cell spacing for a  spectrum and flux consistent with ESS expectations. Solid curves  correspond to metrics using neutrons with $E_n\geq$10\,MeV, dashed curves  to $E_n<$10\,MeV. The opposing trends in event rate and resolution motivate the FOM in the bottom panel (see text).
        A FOM minimum at 40\,cm for the underground corridor  determines the adopted   spacing in the  design.
    }
\end{figure}

Cell spacing was optimized using MCNPX-PoliMi\,\cite{MCNPPoliMi} simulations, aiming to maximize  detection efficiency for an $E_n$ of up to few tens of MeV. Longer center-to-center cell distances improve timing resolution but reduce the probability of interactions involving two cells. A design constraint was to keep  TOF between cells for $E_{n}\sim30$ MeV neutrons at greater than a few ns. The sensitivity to higher $E_n$ decreases rapidly as $E_p$ increases and signal amplitude can surpass the dynamic range of the digitizer (Sec.\ \ref{energy_cal}). Additionally, TOF determination for higher $E_n$ is arduous due to its $\sim\!1$ ns uncertainty (Sec.\ \ref{time}) for which the 500 MS/s sampling rate of the  digitizer is also a limiting factor.

Fig.\ ~\ref{fig:design} summarizes this optimization of  detector cell spacing (cube side length), done in the context of the planned deployment at the ESS, i.e., informed by expected neutron spectra derived from simulations in \cite{lewis,leire}.
The three panels show, as a function of  cell centroid spacing, the expected number of two-cell coincidences  ("events")
from a three-month exposure (top), the average energy and angular resolutions (middle), and a combined figure of merit (FOM, bottom).
Results are shown separately for two example NSC installation sites ---the aforementioned utility rooms 15\,m from the ESS spallation target (red) and an underground corridor at 24\,m (black)--- and for neutrons of energy $E_n\!\!<$10\,MeV and $E_n\!\geq$10\,MeV.
The top panel shows that shorter spacings yield significantly more cell coincidences, as expected.
The middle panel displays the average Gaussian energy resolution $\sigma_E/E_n$ and the relative angular resolution $\sigma_\theta/\theta_n$, which decrease monotonically with increased spacing due to the longer TOF. Reconstruction of  simulated events, smeared in each parameter, is done according to the characterizations in Sec.\ \ref{calibration}.
The opposing trends in event rate  and resolution suggest a combined FOM,
\begin{equation}
    \mathrm{FOM} = \frac{1}{\sqrt{N_{evt}}} + \frac{\sigma_E}{E_n} + \frac{\sigma_\theta}{\theta_n} \quad ,
    \label{eq:fom}
\end{equation}
where $N_{evt}$ is the number of detected events in three months.

A smaller FOM indicates better overall performance: the first term penalizes low statistics, while the second and third terms penalize poor energy and angular resolution, respectively.
Optimal spacing balances these competing contributions. Minimum FOM dictates the preferred design. Keeping in mind the much lower event rate in the better shielded underground corridor and the need for reasonably short exposures while exploring the ESS facility, a cell spacing of 40 cm was settled for.

\section{Calibration} \label{calibration}

\subsection{Energy} \label{energy_cal}

Gain matching of the detector cells was done through a combination of selection of PMT bias and of adjustable gain per channel in the PS 777 linear amplifier (Fig. \ref{fig:daq}). A compromise was sought between sufficient gain for efficient  n-$\gamma$ discrimination (Sec.\ \ref{psd}) and maintaining a 9.5-10 MeV dynamic range per channel, enough for the detection of $E_p\gtrsim$ 20 MeV once proton recoil quenching is considered. At the selected gains no detectable PMT saturation was observed, guaranteeing linearity between integrated amplitude under digitized scintillation pulses ("charge" $Q$) and energy deposited in the cells. 

As for most detector materials, energy depositions by nuclear recoils in organic scintillators are considerably quenched \cite{carbon_qf}. As a result, the detection of neutrons with $E_n$ up to few tens of MeV requires a scintillator energy range of only up to a few MeV.
Two $\gamma$-ray emitters ($^{22}$Na and $^{137}$Cs) were used to calibrate the electron-recoil energy scale of the cells. Compton scattering dominates the gamma response of small-volume organic scintillators.
Gain matching roughly aligned the energy spectra from each cell, ensuring a balanced response.

Compton edges provide sharp and calculable reference features in these calibration spectra. However, the finite energy resolution of the detector smears these edges, complicating their definition. A simple approach is to define edge position at the half-maximum of the Compton continuum, fitting to it a normal distribution ~\cite{chikkur1973new}. Dietze and Klein~\cite{dietze1982gamma} later showed via Monte Carlo simulation that the true Compton edge lies between the maximum of the spectrum and the half-maximum, with the exact value depending on  detector resolution,  size, and the incident $\gamma$-ray energy. More elaborate methods include facing two detectors to exploit backscatter coincidences~\cite{nagvi1991energy,swiderski2010measurement,tajudin2020response}, or fitting the full spectral shape with a Geant4 or MCNP detector model that incorporates the energy resolution~\cite{siciliano2008energy,hohara2001simple,RanjbarKohan2012,ghadiri2015studying,mengesha2017method,mauritzson2022geant4}.

Here we adopt the last of these approaches and use MCNPX-PoliMi 
to simulate the  deposited energy $E$ of $\gamma$-rays in the scintillators. Charge spectra (i.e., integrated pulse amplitude from DAQ waveforms, proportional to the energy deposited in the cell) are compared to these simulations, which are smeared with a 
resolution function of the form
\begin{equation}
    \frac{\mathrm{FWHM}}{E} = \sqrt{\alpha + \frac{\beta}{E}}\,,
    \label{eq:resolution_intro}
\end{equation}
where FWHM is the full width at half maximum of the Gaussian resolution,  $\alpha$ and $\beta$ are energy-independent and stochastic 
contributions, respectively. The linear  
calibration coefficients correlating charge and energy and the two resolution parameters $\alpha$ and $\beta$ are 
determined simultaneously via Bayesian Markov-Chain Monte Carlo using the 
Metropolis--Hastings algorithm \cite{Hastings}.

\begin{figure}
    \centering
    \includegraphics[width=0.9\linewidth]{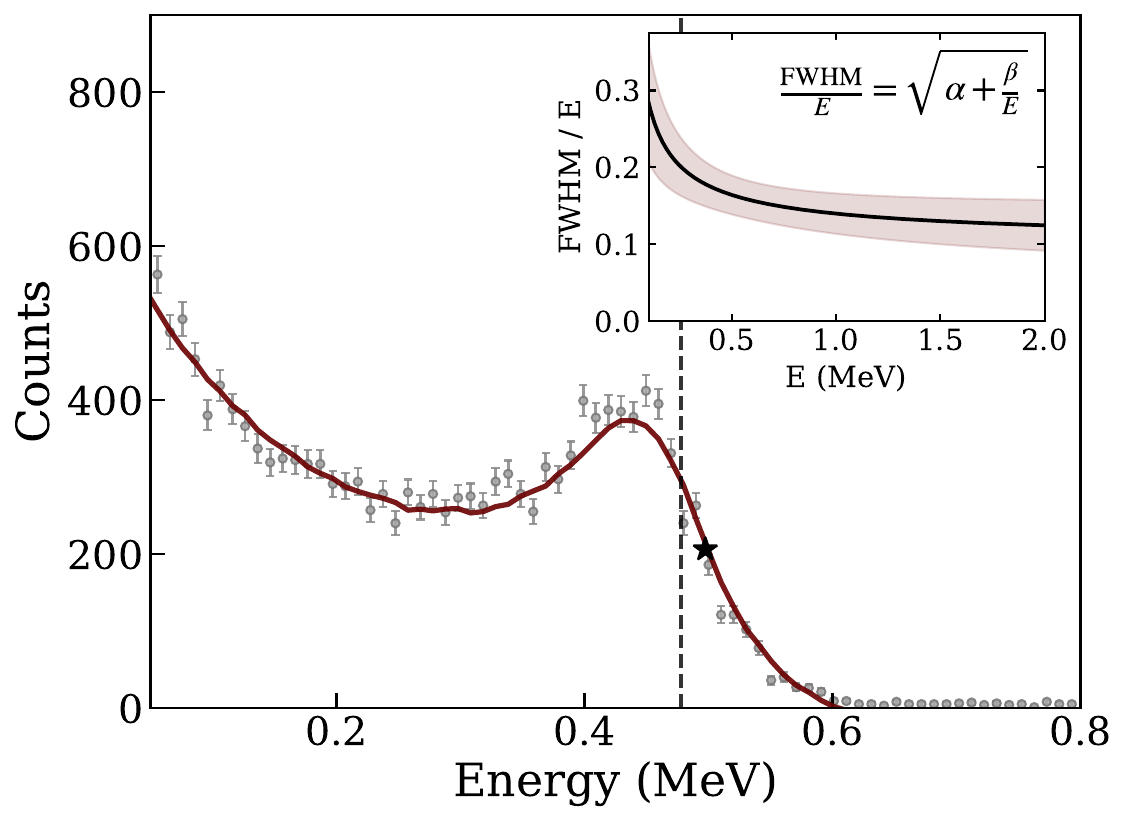}
    \caption{
    Calibration energy spectrum (datapoints) for a representative detector cell exposed to a $^{137}$Cs gamma source. Overlaid is a simulation smeared by the best-fit energy resolution (see text). The calculated Compton edge at 478\,keV is indicated by the dashed vertical line. A star marks the half-maximum point. The inset depicts the mean FWHM/$E$ of all cells and the shaded band the $\pm1\sigma$ variation between them.
     }
    \label{fig:energy_cal}
\end{figure}

A representative $^{137}$Cs calibration spectrum in Fig.\ \ref{fig:energy_cal}  illustrates the good agreement between simulation and measurement and the improvement of the present method over  ~\cite{chikkur1973new,dietze1982gamma}. The inset shows  the energy resolution averaged over all detector cells, which follows the typical $1/\sqrt{E}$ dependence expected from scintillators. The  linear relationship between charge and energy included a similar treatment of  $^{22}$Na runs.

\subsection{Time} \label{time}

An accurate TOF measurement between the two scatter sites is essential to determine $E_{TOF}$ (Eq.~\ref{eq:tof}).
The intrinsic jitter of the digitizer is  negligible in this respect. The primary source of uncertainty is the transit time of electrons through the dynode stages of the PMTs. This consists of two effects, both of which are characterized here with the use of simultaneous 511 keV back-to-back gamma emissions from a $^{22}$Na source placed at the center of as many detector cell pairs as viable, with any onset time differences between coincident events directly reflecting systematics of the system.

\begin{figure}
    \centering
    \includegraphics[width=0.96\linewidth]{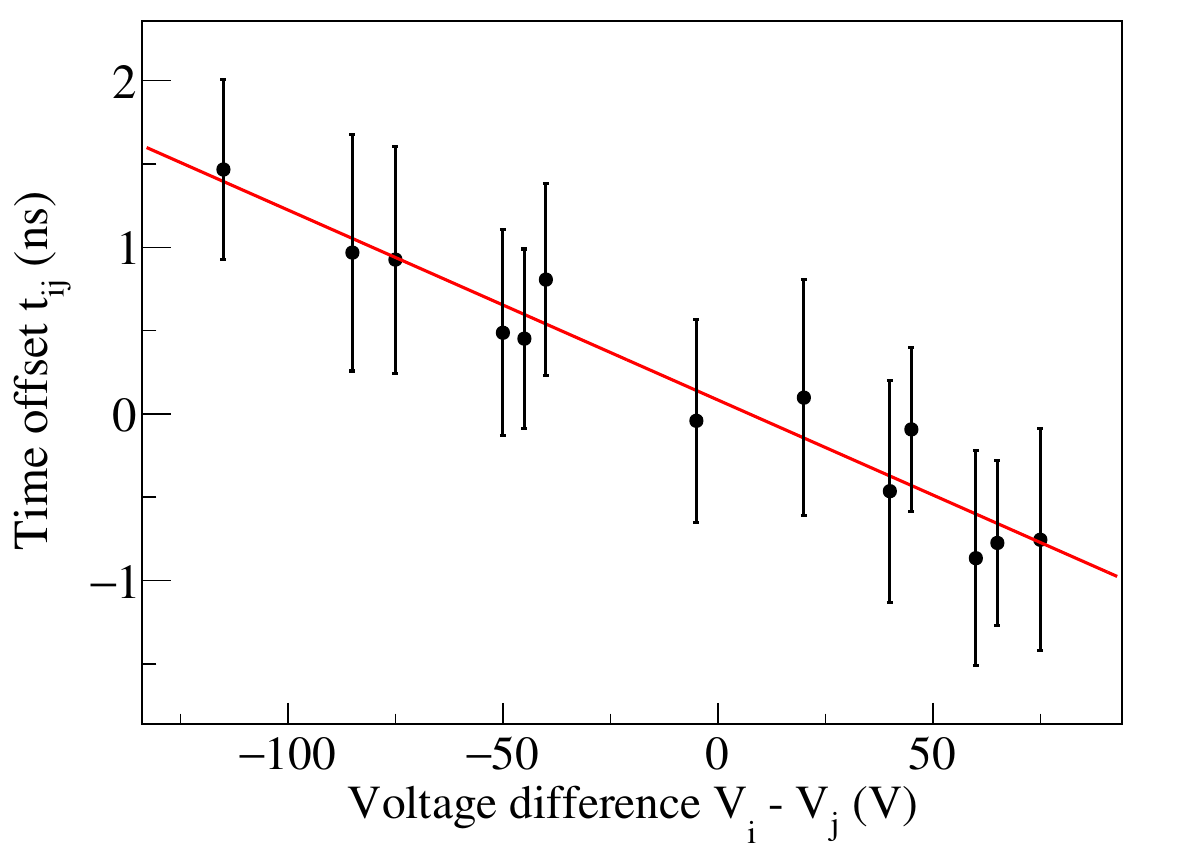}
    \caption{Measured onset-time offset $t_{ij}$ between PMT pairs as a function of the operating voltage difference $V_i - V_j$. The solid line is a linear fit to the data. Error bars represent the standard deviation of the onset-time distribution for each pair, comparable to the TTS of the PMTs (see text).}
    \label{fig:pmt_voltage_offset}
\end{figure}

The first effect is a correctable cell-dependent pulse onset offset that arises from the different PMT operating voltages.
A higher bias supply voltage shortens the electron transit time and steepens the leading edge of  signal pulses, both of which shift the reconstructed onset time~\cite{hamamatsu2017pmt}. Figure~\ref{fig:pmt_voltage_offset} shows the measured onset-time offset as a function of bias difference between PMT pairs for which  $^{22}$Na source positioning in the middle of their cell centroid separation  was easily accomplished. The data are well-described by a linear relationship. The contour describing these offsets was compiled into a correction matrix $t_{ij}$, where $i$ and $j$ denote different PMTs. This matrix is used to adjust onsets event-by-event during the reconstruction of candidate coincidences. Naively this correction could have been bypassed by biasing all PMTs identically and gain-matching them relying exclusively on the variable-gain PS 777 linear amplifier  (Fig. \ref{fig:daq}). Realistically this approach is not always feasible due to the variability in PMT gain across units of the same model and the need for a minimum gain (i.e., bias) to obtain good n-$\gamma$ discrimination (Sec.\ \ref{psd}).

The second effect is a stochastic spread in the onset time of $\sigma \approx 1.2$\,ns (presenting as error bars in Figure~\ref{fig:pmt_voltage_offset}), consistent with the transit-time spread (TTS) of the Hamamatsu H6410 PMT, reported as 1.1\,ns in ~\cite{HamamatsuH6410}. This jitter directly impacts the TOF resolution and therefore the accuracy of the reconstructed scattered-neutron energy and scattering angle. The effect is most pronounced at high neutron energies, where the TOF between cells is short, and a fixed timing uncertainty represents a larger fractional error.

\subsection{Pulse shape n-$\gamma$ discrimination} \label{psd}

Although the pulse shape discrimination (PSD) performance of specialty plastic scintillators is generally lower than that of  liquid equivalents ~\cite{2761,Laplace2020,Shen2022PSD}, the benefits in gained chemical safety and NSC transportability are well worth the penalty for our planned application. 
The PSD parameter was constructed using a customary approach in which the pulse charge $Q$ is integrated over prompt and delayed timescales ($Q_1$ and $Q_2$, respectively) which are optimized in duration to separate neutron-like and gamma-like events \cite{psd1,psd2}.
Datasets with energy depositions purely from gammas obtained using $^{22}$Na and $^{137}$Cs sources were contrasted in $Q_2$ vs.\ $Q_1$ phase space with those from a $^{252}$Cf neutron (and gamma) source. This was done for a variety of integration time choices so as to maximize a figure of merit (FOM) defined as
\begin{equation}
\mathrm{FOM} = \frac{\ell}{\sigma_{n} + \sigma_{\gamma}} \quad ,
\label{eq:FOM}
\end{equation}
where $\ell$ is the separation between centroids of neutron and gamma distributions in PSD parameter space ($Q_2/Q_1$ above 130 keV) and $\sigma_{n,\gamma}$  the standard deviations of those distributions. The optimized integration regions found were 0--60\,ns for $Q_1$ and  60--250\,ns for $Q_2$, where 0 is the onset of the pulse in the waveform.

\begin{figure}
\centering
\includegraphics[width=1.\linewidth]{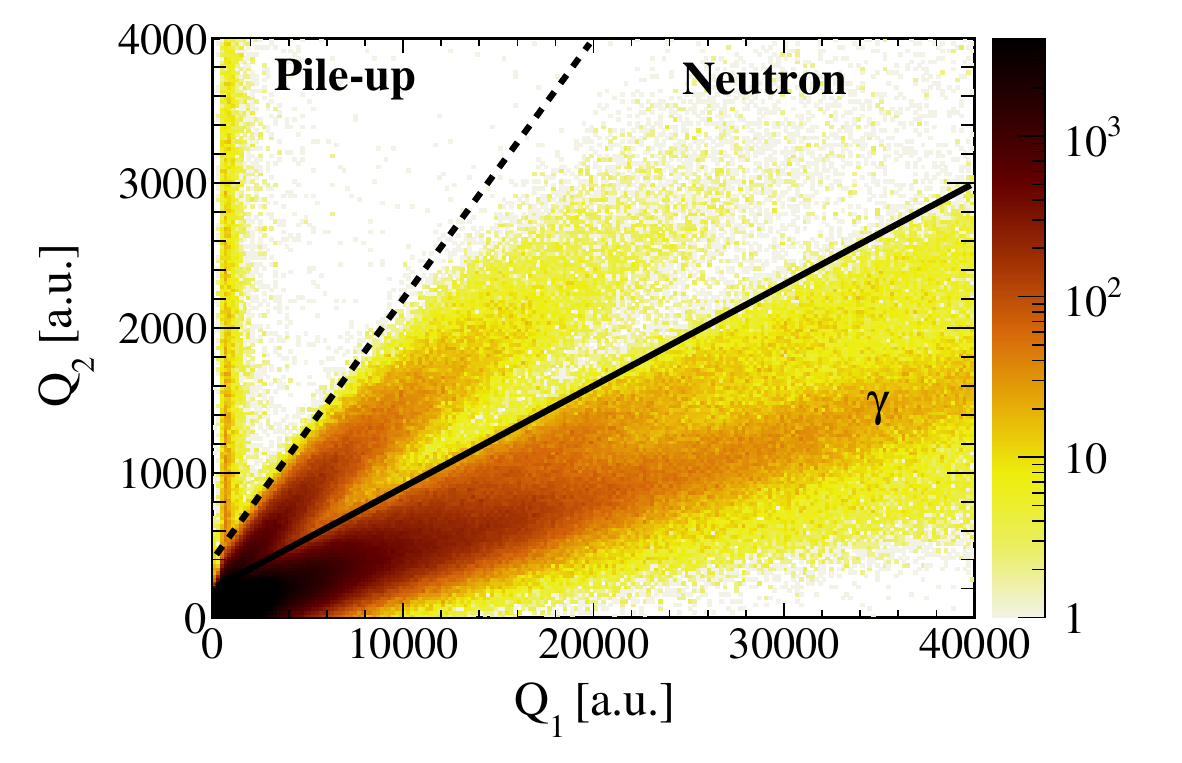}
\caption{$Q_2$ vs.\ $Q_1$ scatter plot for a weak $^{252}$Cf source in direct contact with a detector cell. Gamma-like  and neutron-like populations are visible. Cut boundaries are decided upon with the help of pure gamma sources for which the neutron population is absent. Pile-up events  at anomalously high $Q_2/Q_1$ values are also excluded. Those are mostly absent during NSC normal operation away from intense radiation sources.}
\label{fig:psd_cf}
\end{figure}

The $Q_1$ vs.\ $Q_2$ scatter plot for 
an example $^{252}$Cf dataset is shown in Fig.~\ref{fig:psd_cf}. Only events depositing $>$ 60 keV are  included, as PSD  rapidly loses discrimination power at lower energy. Cuts were educated by the distribution of  gamma-like events from similar datasets taken with the pure gamma sources.
Conservative cuts separating the neutron and gamma populations, shown in the figure, prioritize neutron acceptance at the expense of some gamma contamination. Californium neutrons follow a Maxwellian energy distribution with <$E_n$>$\sim$2.2 MeV. As can be appreciated from Fig.~\ref{fig:psd_cf}, for the typically higher neutron energies expected at spallation sources (up to tens of MeV) PSD improves continuously.

\section{Neutron source reconstruction}

Deployment of the neutron camera will see it accumulate both position and energy information. The only neutron source available for characterization of the device was $^{252}$Cf. Its spontaneous fission neutron spectrum peaks below 2 MeV and produces an average of 3.7 neutrons per fission. Source yield during NSC characterization was  just $\sim850$ n/s, measured with a  $^{3}$He counter surrounded by moderator \cite{drew}. The high neutron multiplicity per fission results in false coincidence events that are as  likely to pass PSD cuts as true coincidences produced by a single neutron. These false coincidences mimic high $E_n$ events. Near-field datasets taken with the source positioned on the aluminum frame of the camera were particularly affected by this effect. Countermeasures are described below.

\subsection{Source energy} \label{energy}

Eqs. \ref{eq:tof} and \ref{eq:total_en} describe the approach to the derivation of incoming neutron energy  $E_n$. In this section we demonstrate a full understanding of the systematics involved by successfully comparing reconstructed and simulated  $^{252}$Cf energy spectra.

\begin{figure}[!tbp]
\centering
\includegraphics[width=.9\linewidth]{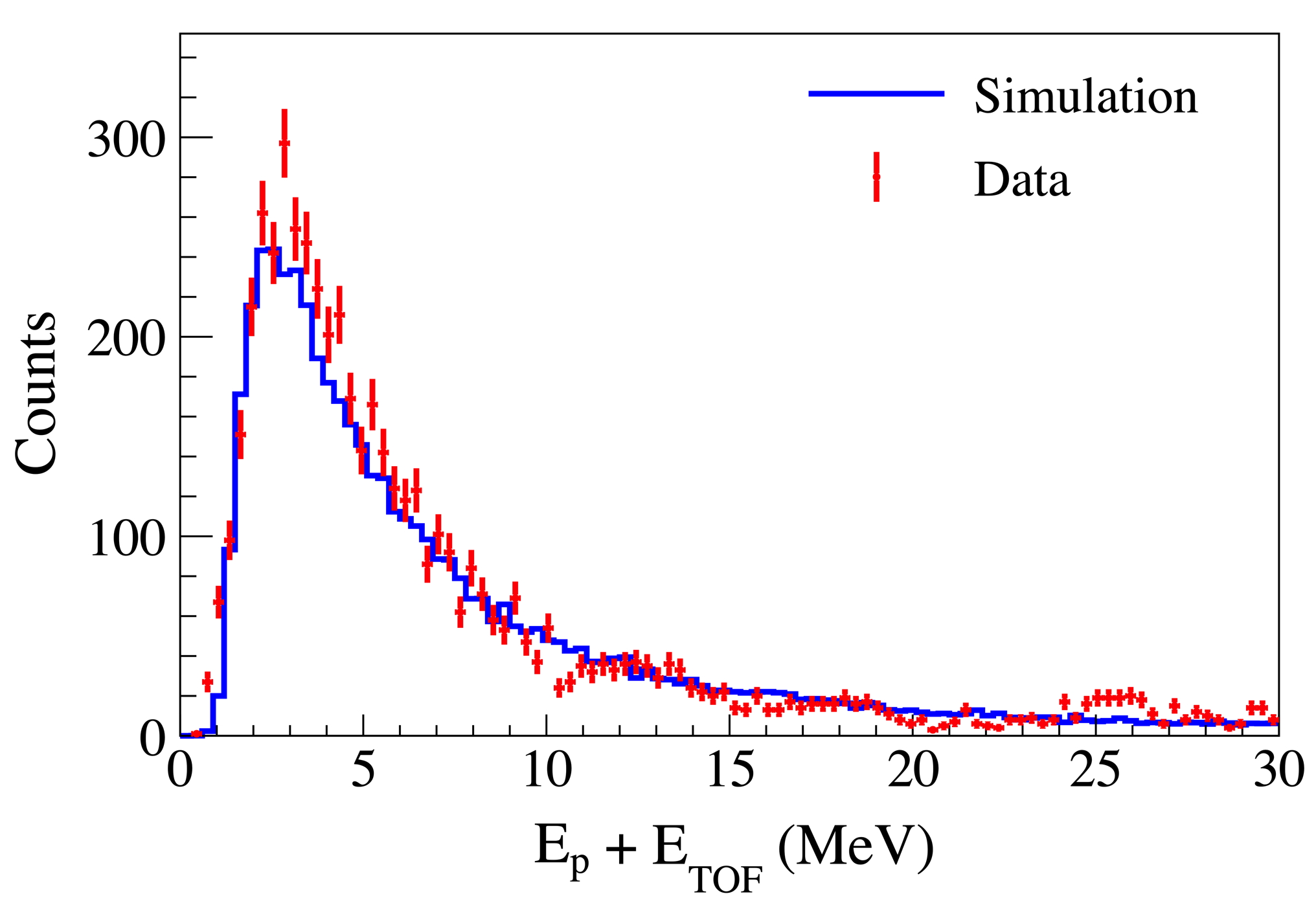}
\caption{\label{fig:energy} {Reconstructed incoming neutron energy ($E_n$, Eq.\ \ref{eq:total_en})  from a weak  $^{252}$Cf source. The solid line represents the simulated spectrum, processed in the same way as the data (see text). $E_p$ is unquenched, i.e., corrected for quenching factor. } }
\end{figure}

The dataset used for this demonstration is the sum of exposures where the $^{252}$Cf source was positioned at the eight corners of the aluminum frame holding detector cells (Fig. \ref{fig:camera}). An energy cut of $E_p>$ 60 keV was imposed to guarantee PSD (Sec.\ \ref{psd}). Similarly, the hardware condition TOF $<$ 120 ns (Sec.\ \ref{design}) restricted $E_{TOF}$ to be above 60-175 keV, depending on $d$ involved in the coincidence. An additional cut (TOF $<$4 ns removed) reduced contamination from $^{252}$Cf fission gammas, which appeared as prompt coincidences.

\begin{figure*}[!ht]
\centering
\includegraphics[width=\textwidth]{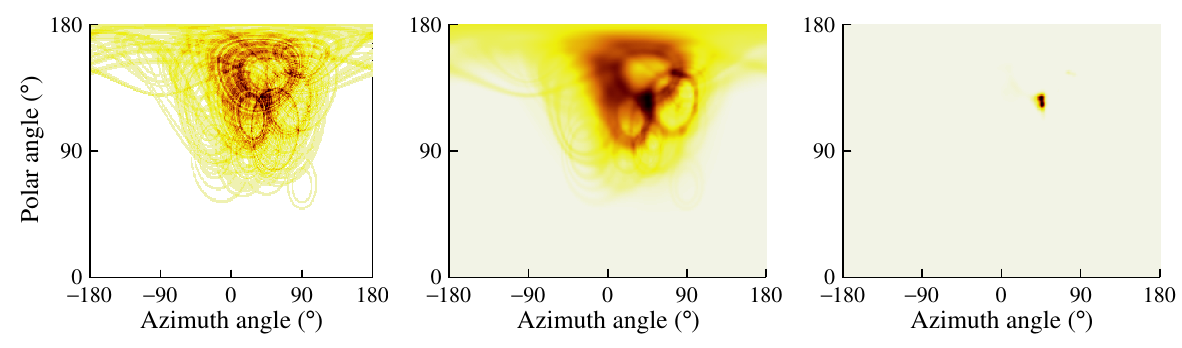}
\caption{\label{fig:heatmap} { Overlap of projected cones in angular space for an example dataset with the neutron source in one corner of the NSC aluminum frame. {\it Left:} Raw back-projected summation of the data. {\it Center:} Summation of the back-projections of each event, smeared by the angular resolution of the camera. {\it Right:} Result of the iterative MLEM algorithm converging on the most likely likelihood distribution from which the neutrons originated (see text). }  }
\end{figure*}

Simulations of the setup were performed with both GEANT4 \cite{geant4} and MCNPX-PoliMi \cite{MCNPPoliMi}. These included source structure (i.e., self-shielding), $^{252}$Cf-specific neutron energy spectrum and the possibility of instantaneous multiple neutron emission. Quenching of energy depositions in the cells followed the models of \cite{ficenec} and \cite{carbon_qf} for proton and carbon recoils, respectively. $E_p$ was smeared with the resolution  in Sec.\ \ref{energy_cal}. The simulated spread of signal onsets  was estimated by combining the PMT characteristics discussed in section \ref{time} with the DAQ timing resolution (1 ns), summed in quadrature. Cell center-to-center distances  are used in TOF calculations rather than simulated interaction positions. The  $\Delta d = 2$\,cm dispersion involved is treated as a geometric uncertainty on  $d$ when using  Eq.~\ref{eq:tof}. These uncertainties, propagated through Eq.~\ref{eq:total_en}, give the per-event uncertainty on $E_n$ that smears the simulated spectrum. The same  propagated through Eq.~\ref{eq:angle_recon} yield the per-event angular uncertainty used for the directional reconstruction in the following section.

Fig.\ \ref{fig:energy} compares the reconstructed neutron-energy distribution in  calibration data with the corresponding simulated response. This comparison should be understood as a validation of the NSC ability to recover the  hardness of incident neutrons, rather than as a deconvolution of the underlying fission spectrum, which we reserve for future work (Sec.\ \ref{comment}). The good agreement between the two indicates that treatment of the data as described is broadly consistent with the NSC response model. Small residual differences between data and simulation arise from a lack of  modeling of structures in the vicinity of the source (room wall and floor) and unmodeled neutron-gamma discrimination criteria (Sec.\ \ref{psd}). Fig. \ref{fig:energy} compares favorably with earlier attempts at $E_n$ reconstruction using  NSCs \cite{nc3,nc4,nc5}.

\subsection{Source position}  \label{position}

\subsubsection{Methodology}

Section \ref{how} describes the limited reconstruction ability for  source origin that a single event provides: a cone in space, with an associated uncertainty. Summing the projections of many such cones on a reference frame  of polar (zenith) and azimuthal coordinates returns an image of the neutron source following sufficient exposure. The simplest approach assigns a binary weight, unity to each angular bin overlapped by a cone, zero elsewhere. However, this neglects the broad likelihood distribution that  finite energy and position resolutions actually impose. The iterative method described in this section respects and profits from that likelihood interpretation.

All event origins are spatially projected assuming elastic scattering off hydrogen occurred. Although carbon scatters constitute roughly 40\% of elastic interactions in PVT-based plastic scintillator, kinematics limits the maximum carbon recoil to $E_{C,\max}  \approx 0.28\,E_n$. Additionally, the carbon quenching factor in organic scintillator falls below 1\% for $E_C>$ 1\,MeV, never surpassing 5\% at lower energy~\cite{carbon_qf}. For  tens-of-MeV neutrons dominant at spallation sources and our $E_p$ > 60\,keV analysis threshold, accepted events are therefore overwhelmingly proton recoils, and the assumption made introduces only a small directional bias.

\begin{figure*}[!t]
\centering
\includegraphics[width=\textwidth]{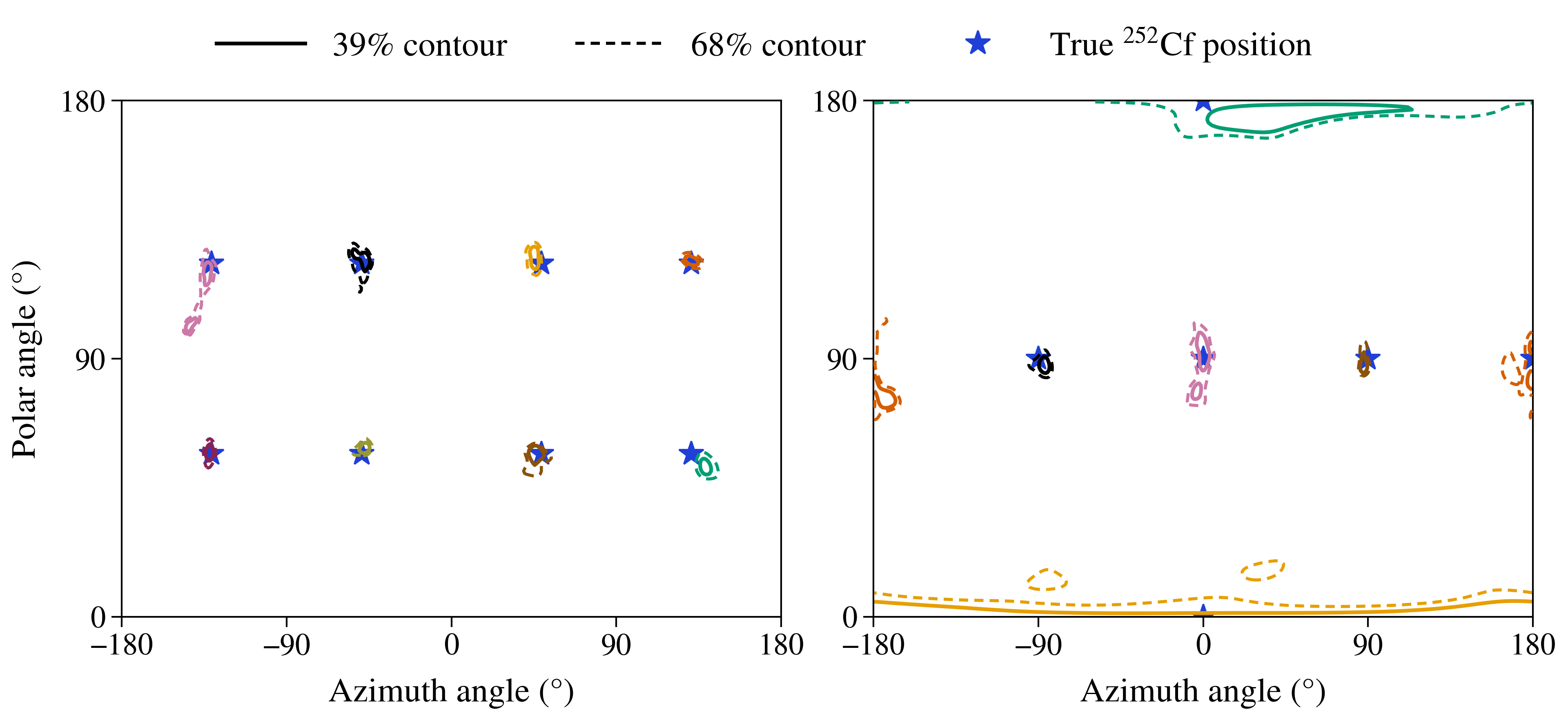}
\caption{\label{fig:4pi} { Angular reconstruction of  datasets covering the full field of view of the neutron camera. {\it Left:} Datasets with the $^{252}$Cf source placed on the eight corners of the cubic aluminum extrusion frame. {\it Right:} Datasets with the source placed at the center of  the six faces of the frame. All exposures were comparable at $\sim$40 h. }  }
\end{figure*}

Fig. \ref{fig:heatmap} illustrates  the steps taken to obtain directional information. The dataset used has the $^{252}$Cf source placed in a corner of the NSC aluminum extrusion frame holding cells. The left panel in the figure shows the binary-weight summed projection. The central panel weights each projection, before summation, by the uncertainty on its reconstructed scattering angle via Gaussian smearing. This reflects that each event is a likelihood distribution rather than a localized binary constraint. The per-event angular uncertainty $\sigma_\theta$ used in this smearing is obtained by propagating the same measurement uncertainty budget delineated in Section~\ref{energy} ($E_p$ and TOF resolution and $\Delta d = 2$\,cm dispersion) through Eq.~\ref{eq:angle_recon}, yielding a typical $\sigma_\theta \approx 3.3^\circ$.

The right panel takes that smeared back-projection and follows an iterative method (maximum likelihood expectation maximization, MLEM) detailed in \cite{Steinberger_2020} and \cite{Hou_2022}, to converge on the most likely angular distribution from which neutrons originate. MLEM emphasizes persistent regions of overlap and reduces pixel-scale fluctuations. The smoothing process is stopped when the variance in image differences between iterations falls below 10\% of the initial change. This  procedure is used for all directional reconstructions in this work.

\subsubsection{$4\pi$ sensitivity}

The defining feature of this NSC is its ability to reconstruct source direction over a 4$\pi$ sphere simultaneously, without orientation toward a presumed source.
To demonstrate this, the $^{252}$Cf source was placed in various locations around the camera, including positions near each corner of the aluminum extrusion frame and  also centered on each of its cubic faces, using a removable holder. These measurements confirm that the trigger logic and cubic design described in Section \ref{design} provide an improvement over the preferentially forward-backward  field of view typical of two-plane NSCs.

For sources distant relative to the magnitude of  $d$, typical of NSC use in the field, the two cells involved in an event are effectively at a common origin, and each event can be represented as a unit vector along its cone axis. Spatial reconstruction is then purely angular and any residual parallax falls below the camera's angular resolution. Present datasets, however, have the  $^{252}$Cf source at distances comparable to $d$, so the dependence of the cone-axis endpoint on the off-center cell positions (i.e., parallax) is non-negligible. To absorb parallax in this atypical near-field calibration mode, imposed by the low source yield, cones are projected onto a sphere of variable radius, chosen to maximize the overlap of the summed smeared likelihood distributions, in a distance-estimating procedure akin to that in \cite{Hou_2022}.

Fig.\ \ref{fig:4pi} summarizes the reconstructed source positions for all datasets, compared to  known location. Overlap between true and reconstructed positions is affected by spurious event reconstructions arising from multiple neutron emission in $^{252}$Cf. To suppress this effect events with TOF\,$<$\,6\,ns are removed. This threshold  is defined by a simulation of $^{252}$Cf neutrons that targets the false-coincidence population generated by the 3.7 average multiplicity in neutron emission. Reconstructions that violate the kinematic limit of neutron--proton elastic scattering (i.e., a reconstructed scattering angle $>90^\circ$) are likewise rejected. Following an accumulation of 480 events on average after cuts, the reconstructed hotspot agrees with true source location within $\sim3^\circ$ in both polar and azimuthal angles for all fourteen locations tested, after comparable ($\sim$40 h) exposures. This  agreement demonstrates that the NSC and the reconstruction pipeline  recover the source directions robustly and uniformly across the full spatial 4$\pi$ field of view.

\subsubsection{Overlay on panoramic camera images}

Although near-field $^{252}$Cf measurements are useful for validating the isotropic response of the instrument, the intended deployment at spallation facilities will involve sources located at greater distances. To probe this regime the source was placed  $1.0$--$1.5$\,m away from the camera for longer exposures (8-35 days), necessary due to the weak 850 n/s yield. 

\begin{figure}
\centering
\includegraphics[width=1\linewidth]{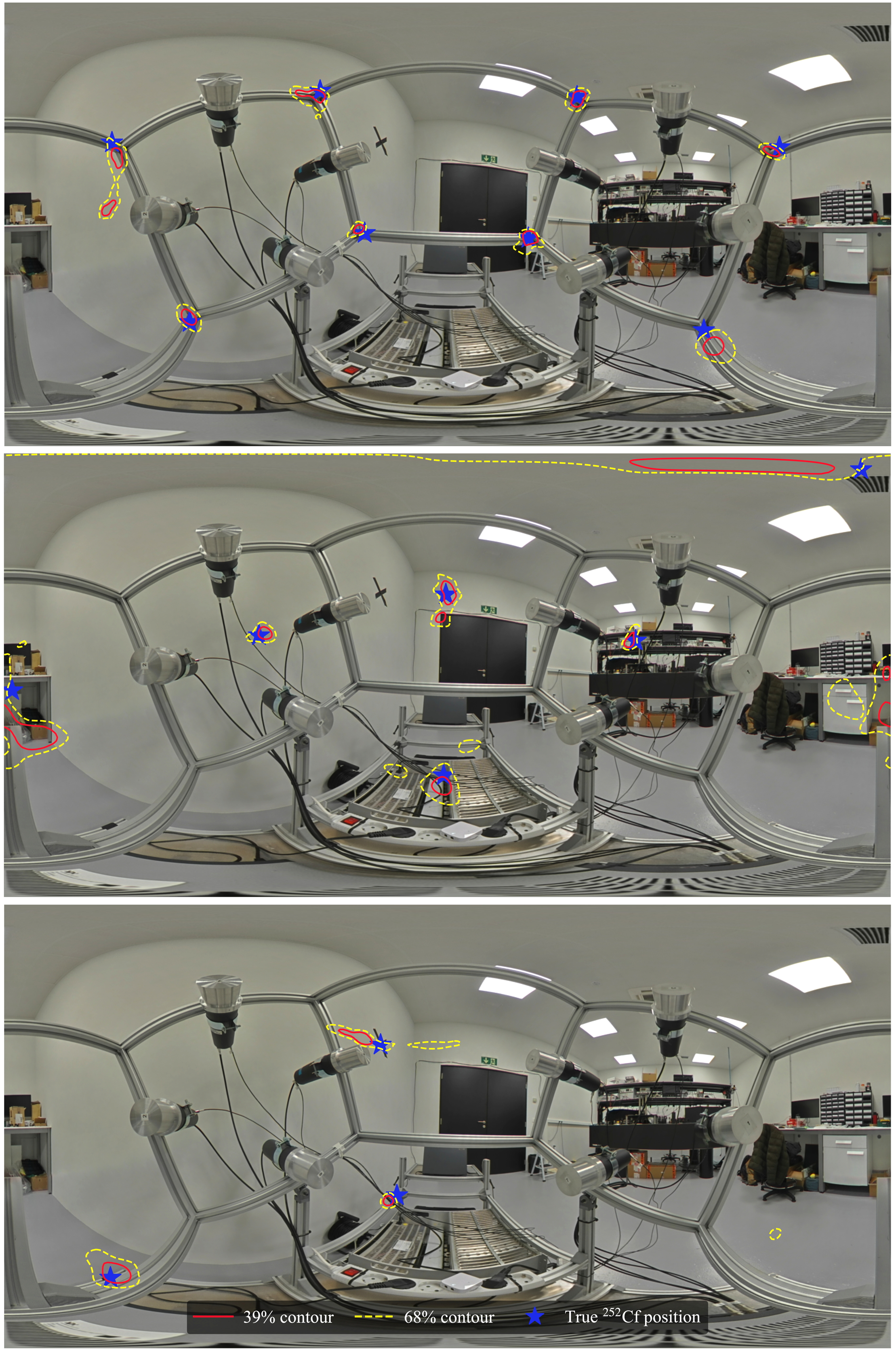}
\caption{\label{fig:overlay} { Angular reconstruction of $^{252}$Cf measurements overlaid on a panoramic image from the built-in 360$^{\circ}$ camera. The true position of the source in each dataset is marked by a blue star. Top and middle panels correspond to measurements in Fig.\ \ref{fig:4pi}, bottom panel to three far-field tests (see text). During middle panel runs the source was placed at the center of each aluminum frame face with a removable holder, absent in this image.}  }
\end{figure}

A 360$^{\circ}$ panoramic camera (Ricoh Theta SC2) is an integral part of the NSC (Fig.\ \ref{fig:camera}). Its off-center positioning is intentional, in order  not to encumber neutron trajectories between cells, otherwise free of impediment. This 360$^{\circ}$ camera provides a seamless equirectangular projection of ($\theta$,$\phi$) coordinates onto pixel count along (x,y) 2D images. Internal tilt sensors automatically center its images along pitch and roll. This facilitates the projection of reconstructed neutron source positions onto a visual representation of the environment around the NSC (Fig.~\ref{fig:overlay}), once the panoramic camera off-center displacement is accounted for. A third degree of freedom (yaw) is not automatically corrected for and can be altered via a thumbscrew at the base of the 360$^{\circ}$ camera, used to secure its position, coupling it to the  NSC. We ascertained that a best-effort to make 360$^{\circ}$ camera images L-R symmetric as in Fig.~\ref{fig:overlay} is sufficient to render the effect of precise yaw positioning negligible. This was done by mounting panels containing a grid of positions on all sides of the aluminum frame. Their nominal angular positions coincided with those from panoramic camera images  assuming  yaw centering in the equirectangular projection, within $\sim\! 1.5^{\circ}$.

\section{Conclusions} \label{comment}

The far-field measurements described in the previous section allow us to quantify some of the performance expectations for this NSC. The run corresponding to $^{252}$Cf positioning 1.5 m away on a nearby wall, marked with a black tape cross and visible on Fig.~\ref{fig:overlay} corresponded to an exposure of 12.5 days, resulting in an accumulated 209 events passing cuts, i.e., an event rate over 3 months of  $N_{evt}=$ 1490 (Fig.\ ~\ref{fig:design}, Eq.\ \ref{eq:fom}). The weak neutron yield of the source resulted in a flux at the NSC position (cube center) of just $\sim 0.003$ n/cm$^{2}$s. This is comparable to the $\sim 0.0024$ n/cm$^{2}$s  diffuse flux of 0.1 MeV--20 MeV environmental neutrons at sea level \cite{HU201978}, indicating a good sensitivity to low neutron fluxes with modest signal-to-background ratio, within reasonable time-scales. 

Additionally, the $E_n<10$ MeV event rates $N_{evt}$ for cell spacing of 40 cm shown in the top panel of Fig.\ ~\ref{fig:design} correspond to simulated fluxes of $\sim$5 n/cm$^{2}$s (utility room) and $\sim$0.05 n/cm$^{2}$s (underground corridor), which are expected from full-power \mbox{(5 MW)} ESS operation \cite{lewis}. Those rates and fluxes scale correctly with respect to the same for the 1.5 m $^{252}$Cf run (see values above). This provides an indirect validation for the simulations in Fig.\ ~\ref{fig:design}. Moreover, the quality of the reconstruction already obtained with just 209 $^{252}$Cf  events indicates that short runs (few hours or days) will suffice for ESS studies at most locations, even when initially operating at a reduced power. In this respect it should be emphasized that most beam-related neutrons at spallation facilities arrive within a few $\mu$s following their production at the target. Both the hardware and internal logic triggers discussed in Sec.\ \ref{design} can be trivially gated \footnote{In the first case using the veto input of either PS 710 discriminator, in the second via the GPI input to the digitizer.}. to be ready only during those pulsed periods,  decreasing NSC sensitivity to backgrounds unrelated to POT. 

The  portable NSC presented here is a clear candidate for early deployment at the ESS, coinciding with periods of initial optimization of neutron shielding around planned instruments and detectors, serving the facility and its community. Its readiness and desired properties have been demonstrated in this work. Nevertheless, future improvements can be envisioned. For instance, the NSC  decreasing efficiency outside of 0.2 MeV $\lesssim E_n \lesssim$ 30 MeV should be carefully mapped via additional simulations, so as to allow for an accurate deconvolution of source energy spectrum and flux. It should also be possible to improve the NSC user interface to provide real-time progressive updates on  source characteristics (location, hardness and flux for each individual source present) even for complex neutron fields such as those expected within ESS instrument caves.

\section*{CRediT authorship contribution statement}
\textbf{S.G. Yoon:} Methodology, Software, Validation, Formal analysis, Investigation, Writing -- original draft, Visualization.
\textbf{L. Larizgoitia:} Methodology, Software, Validation, Formal analysis, Investigation, Writing -- original draft, Visualization. \textbf{C.M. Lewis:} Methodology, Software, Investigation, Writing -- original draft, Supervision. \textbf{F. Monrabal:} Writing -- original draft, Supervision, Funding acquisition. \textbf{J.I. Collar:} Conceptualization, Methodology, Investigation, Resources, Writing -- review \& editing, Supervision, Funding acquisition.

\section*{Declaration of competing interest}
The authors declare that they have no known competing financial
interests or personal relationships that could have appeared to
influence the work reported in this paper.

\section*{Acknowledgements}
This work is supported by ERC Advanced Grant 101055120 (ESSCEvNS), ERC Starting Grant 
101039048 (GanESS) and partially by NSF Award PHY-2209579. L.L. acknowledges support from a Spain–U.S. Fulbright Predoctoral Fellowship. We gratefully acknowledge J. Cederk{\"a}ll, E. Lytken and V. Santoro for their continuous encouragement of this work. 

\section*{Data availability}
Data will be made available on request.

\bibliographystyle{elsarticle-num}
\bibliography{bibliography}

@article{MCNPPoliMi,
  title={{MCNP-PoliMi}: a {Monte-Carlo} code for correlation measurements},
  author={Pozzi, Sara A and Padovani, Enrico and Marseguerra, Marzio},
  journal={Nucl. Instrum. Meth. A},
  volume={513},
  number={3},
  pages={550--558},
  year={2003},
  publisher={Elsevier}
}

@article{geant4,
  title={{GEANT4}---a simulation toolkit},
  author={Agostinelli, S. and others},
  journal={Nucl. Instrum. Meth. A},
  volume={506},
  number={3},
  pages={250--303},
  year={2003},
  doi={10.1016/S0168-9002(03)01368-8}
}

@phdthesis{lewis,
    author = "Lewis, C.M.",
    title = "{Particle physics in the sub-keV energy regime}",
    eprint = "2310.01314",
    archivePrefix = "arXiv",
    primaryClass = "physics.ins-det",
    doi = "10.6082/uchicago.5730",
    school = "U. of Chicago",
    year = "2023"
}

@article{siciliano2008energy,
  title = {Energy calibration of gamma spectra in plastic scintillators using {Compton} kinematics},
  author = {Siciliano, E. R. and Ely, J. H. and Kouzes, R. T. and Schweppe, J. E. and Strachan, D. M. and Yokuda, S. T.},
  journal = {Nucl. Instrum. Meth. A},
  volume = {594},
  number = {2},
  pages = {232--243},
  year = {2008},
  publisher = {Elsevier},
  doi = {10.1016/j.nima.2008.06.031}
}

@article{dietze1982gamma,
  title={Gamma-calibration of {NE} 213 scintillation counters},
  author={Dietze, G. and Klein, H.},
  journal={Nucl. Instrum. Meth.},
  volume={193},
  number={3},
  pages={549--556},
  year={1982},
  publisher={Elsevier},
  doi={10.1016/0029-554X(82)90249-X}
}

@article{chikkur1973new,
  title = {A new method of determining the {Compton} edge in liquid scintillators},
  author = {Chikkur, G. C. and Umakantha, N.},
  journal = {Nucl. Instrum. Meth.},
  volume = {107},
  number = {1},
  pages = {201--202},
  year = {1973},
  publisher = {North-Holland Publishing Co.},
  doi = {10.1016/0029-554X(73)90034-7},
  url = {https://www.sciencedirect.com/science/article/abs/pii/0029554X73900347}
}

@article{swiderski2010measurement,
  title = {Measurement of {Compton} edge position in low-{Z} scintillators},
  author = {Swiderski, Lukasz and Moszy{\'n}ski, Marek and Czarnacki, Wies{\l}aw and Iwanowska, Joanna and Syntfeld-Ka{\.z}uch, Agnieszka and Szcz{\k{e}}{\'s}niak, Tomasz and Pausch, Guntram and Plettner, Cristina and Roemer, Katja},
  journal = {Radiat. Meas.},
  volume = {45},
  number = {3-6},
  pages = {605--607},
  year = {2010},
  publisher = {Elsevier},
  doi = {10.1016/j.radmeas.2009.10.015},
  url = {https://doi.org/10.1016/j.radmeas.2009.10.015}
}

@article{nagvi1991energy,
  title = {Energy resolution tests of 125 mm diameter cylindrical {NE213} detector using monoenergetic gamma rays},
  author = {Nagvi, A. A. and Al-Juwair, H. and Gul, K.},
  journal = {Nucl. Instrum. Meth. A},
  volume = {306},
  number = {1-2},
  pages = {267--271},
  year = {1991},
  publisher = {Elsevier},
  doi = {10.1016/0168-9002(91)90331-J}
}

@article{tajudin2020response,
  title={Response of plastic scintillator to gamma sources},
  author={Tajudin, Suffian M. and Namito, Y. and Sanami, T. and Hirayama, H.},
  journal={Appl. Radiat. Isot.},
  volume={159},
  pages={109086},
  year={2020},
  publisher={Elsevier},
  doi={10.1016/j.apradiso.2020.109086},
  issn={0969-8043}
}

@article{mauritzson2022geant4,
  title = {{GEANT4}-based calibration of an organic liquid scintillator},
  author = {Mauritzson, N. and Fissum, K.G. and Perrey, H. and Annand, J.R.M. and Frost, R.J.W. and Hall-Wilton, R. and Al Jebali, R. and Kanaki, K. and Maulerova-Subert, V. and Messi, F. and Rofors, E.},
  journal = {Nucl. Instrum. Meth. A},
  volume = {1023},
  pages = {165962},
  year = {2022},
  issn = {0168-9002},
  doi = {10.1016/j.nima.2021.165962},
  url = {https://doi.org/10.1016/j.nima.2021.165962},
  publisher = {Elsevier}
}

@article{hohara2001simple,
  title={A Simple Method of Energy Calibration for Thin Plastic Scintillator},
  author={Hohara, S. and Saiho, F. and Tanaka, J. and Aoki, S. and Uozumi, Y. and Matoba, M.},
  journal={IEEE Trans. Nucl. Sci.},
  volume={48},
  number={4},
  pages={1172--1176},
  year={2001},
  month={Aug},
  publisher={IEEE},
  doi={10.1109/23.958745}
}

@article{RanjbarKohan2012,
  title = {Modelling plastic scintillator response to gamma rays using light transport incorporated {FLUKA} code},
  author = {Ranjbar Kohan, M. and Etaati, G. R. and Ghal-Eh, N. and Safari, M. J. and Afarideh, H. and Asadi, E.},
  journal = {Appl. Radiat. Isot.},
  volume = {70},
  number = {4},
  pages = {864--867},
  year = {2012},
  issn = {0969-8043},
  doi = {10.1016/j.apradiso.2012.01.014},
  publisher = {Elsevier}
}

@article{ghadiri2015studying,
  title={Studying the response of a plastic scintillator to gamma rays using the {Geant4} {Monte Carlo} code},
  author={Ghadiri, Rasoul and Khorsandi, Jamshid},
  journal={Appl. Radiat. Isot.},
  volume={99},
  pages={63--68},
  year={2015},
  publisher={Elsevier},
  doi={10.1016/j.apradiso.2015.02.017},
  url={https://doi.org/10.1016/j.apradiso.2015.02.017}
}

@article{mengesha2017method,
  title={A method for calibrating the relative gamma-ray light yield of plastic scintillators},
  author={Mengesha, W. and Feng, P. L. and Cordaro, J. G. and M. R. Anstey and Myllenbeck, N. R. and Throckmorton, D. J.},
  journal={Rev. Sci. Instrum.},
  volume={88},
  number={3},
  pages={035108},
  year={2017},
  publisher={AIP Publishing},
  doi={10.1063/1.4978288},
  url={https://doi.org/10.1063/1.4978288}
}

@manual{hamamatsu2017pmt,
  title        = {Photomultiplier Tubes: Basics and Applications},
  author       = {{Hamamatsu Photonics K.K.}},
  edition      = {4th},
  year         = {2017},
  organization = {Hamamatsu Photonics K.K.},
  address      = {Hamamatsu City, Japan},
  url          = {https://www.hamamatsu.com/content/dam/hamamatsu-photonics/sites/documents/99_SALES_LIBRARY/etd/PMT_handbook_v4E.pdf},
  note         = {Accessed: 2026-02-26}
}

@manual{HamamatsuH6410,
  title        = {Photomultiplier Tube Assembly H6410},
  author       = {{Hamamatsu Photonics K.K.}},
  year         = {2023},
  note         = {Data sheet},
  url          = {https://www.hamamatsu.com/jp/en/product/type/H6410/index.html},
  urldate      = {2026-02-26}
}

@article{Laplace2020,
  title = {Comparative scintillation performance of {EJ}-309, {EJ}-276, and a novel organic glass},
  author = {Laplace, T. A. and Goldblum, B. L. and Bevins, J. E. and {Bleuel}, D. L. and {Bourret}, E. and {Brown}, J. A. and {Callaghan}, E. J. and {Carlson}, J. S. and {Feng}, P. L. and {Gabella}, G. and {Harrig}, K. P. and {Manfredi}, J. J. and {Moore}, C. and {Moretti}, F. and {Shinner}, M. and {Sweet}, A. and {Sweger}, Z. W.},
  journal = {JINST},
  volume = {15},
  number = {11},
  pages = {P11020},
  year = {2020},
  month = {nov},
  doi = {10.1088/1748-0221/15/11/P11020},
  publisher = {IOP Publishing}
}

@article{Shen2022PSD,
  title = {{PSD} performance of {EJ}-276 and {EJ}-301 scintillator readout with {SiPM} array},
  author = {Shen, Fengzhao and Pan, Yongyu and Fu, Qibin and Lin, Shaopeng and Huang, Tuchen and Wang, Wei},
  journal = {Nucl. Instrum. Meth. A},
  volume = {1039},
  pages = {167148},
  year = {2022},
  publisher = {Elsevier},
  doi = {10.1016/j.nima.2022.167148},
  issn = {0168-9002}
}

@article{carbon_qf, title={Light output response of {KamLAND} liquid scintillator for protons and $^{12}${C} nuclei}, volume={622}, ISSN={01689002}, DOI={10.1016/j.nima.2010.07.087}, abstractNote={The light output responses for protons and carbon nuclei in the KamLAND liquid scintillator were precisely measured using a monochromatic neutron beam. The observed response for proton recoils is well described by Birks formula with the higher-order correction, for the recoil energy from 424 keV to 10.5 MeV. The response for carbon recoils is also well described by modifying the Birks formula with the nuclear energy loss terms, for the measured recoil energy from 171 keV to 2.2 MeV. The obtained light output responses were used to estimate the background energy spectrum of the 13C(a,n)16O reaction in the KamLAND. The systematic uncertainty of the energy scale in the expected background spectrum was further improved with the measured light output responses.}, number={3}, journal={Nucl. Instrum. Meth. A}, author={Yoshida, S. and others}, year={2010}, month=oct, pages={574--582}, language={en} }

@article{Hastings,
  title={{Monte Carlo} sampling methods using {Markov} chains and their applications},
  author={Hastings, W. K.},
  journal={Biometrika},
  year={1970},
  volume={57},
  number={1},
  pages={97--109},
  doi={10.1093/biomet/57.1.97}
}

@article{Steinberger_2020, title={Imaging Special Nuclear Material using a Handheld Dual Particle Imager}, volume={10}, rights={2020 The Author(s)}, ISSN={2045-2322}, DOI={10.1038/s41598-020-58857-z}, abstractNote={A compact radiation imaging system capable of detecting, localizing, and characterizing special nuclear material (e.g. highly-enriched uranium, plutonium\ldots) would be useful for national security missions involving inspection, emergency response, or war-fighters. Previously-designed radiation imaging systems have been large and bulky with significant portions of volume occupied by photomultiplier tubes (PMTs). The prototype imaging system presented here uses silicon photomultipliers (SiPMs) in place of PMTs because SiPMs are much more compact and operate at low power and voltage. The SiPMs are coupled to the ends of eight stilbene organic scintillators, which have an overall volume of 5.74 $\times$ 5.74 $\times$ 7.11 cm3. The prototype dual-particle imager's capabilities were evaluated by performing measurements with a 252Cf source, a sphere of 4.5 kg of alpha-phase weapons-grade plutonium known as the BeRP ball, a 6 kg sphere of neptunium, and a canister of 3.4 kg of plutonium oxide (7% 240Pu and 93% 239Pu). These measurements demonstrate neutron spectroscopic capabilities, a neutron image resolution for a Watt spectrum of 9.65 $\pm$ 0.94$^\circ$ in the azimuthal direction and 22.59 $\pm$ 5.81$^\circ$ in the altitude direction, imaging of gamma rays using organic scintillators, and imaging of multiple sources in the same field of view.}, number={1}, journal={Scientific Reports}, publisher={Nature Publishing Group}, author={Steinberger, William M. and Ruch, Marc L. and Giha, Nathan and Fulvio, Angela Di and Marleau, Peter and Clarke, Shaun D. and Pozzi, Sara A.}, year={2020}, month=feb, pages={1855}, language={en} }

@article{Hou_2022, title={A neutron scatter imaging technique with distance determining capability}, volume={1022}, ISSN={01689002}, DOI={10.1016/j.nima.2021.165975}, abstractNote={Conventionally, a neutron scatter camera was used to determine the direction of neutron sources. Therefore, various investigations have been conducted on the design of cameras and imaging algorithms. However, the source distance cannot be determined using the existing neutron cameras. In this study, inspired by binocular vision, an approach was proposed to locate the neutron source (both direction and distance) based on a neutron scatter camera composed of two imagers. Each imager reconstructed an image. The source distance was estimated by comparing the two images. An intersection-based back projection imaging (IBPI) method is proposed to suppress the appearance of fake high-luminosity regions in conventional back-projection imaging (CBPI). An imager with four liquid scintillator detectors (one front and three back) was constructed to test the locating approach at the State Nuclear Security Technology Center (SNSTC). A 252Cf isotope source of 3.0 $\times$ 107 Bq was employed as the neutron source in the experiment. Gamma rays and neutrons were discriminated using the pulse shape discrimination (PSD) method. The angular resolution in the experiment was improved from 25$^\circ$ to 13$^\circ$ using IBPI . The binocular distance measurement (BDM) method was successfully applied to estimate the source distance. The experimental results indicate that the source distance is determined more unambiguously using IBPI+BDM than CBPI+BDM. A Monte Carlo simulation was also performed to investigate the design of such a scatter imaging system of the neutron. A relatively large separation between the two imagers was crucial for estimating the source distance. When changing the separation between the two imagers from 40 to 160 cm, FWHM of the estimated source distance could be improved from 9.5 to 4.1 cm.}, journal={Nucl. Instrum. Meth. A}, author={Hou, Yingwei and Song, Yushou and Hu, Liyuan and Liu, Huilan and Zhou, Zhibo and Wu, Zhaohui}, year={2022}, month=jan, pages={165975}, language={en} }

@INPROCEEDINGS{nc1,
  author={Brennan, James and Cooper, Robert and Gerling, Mark and Marleau, Peter and Mascarenhas, Nick and Mrowka, Stanley},
  booktitle={IEEE Nuclear Science Symposium \& Medical Imaging Conference}, 
  title={Applying the neutron scatter camera to treaty verification and warhead monitoring}, 
  year={2010},
  volume={},
  number={},
  pages={691-694},
  doi={10.1109/NSSMIC.2010.5873848}}

@ARTICLE{nc2,
  author={Mascarenhas, Nicholas and Brennan, James and Krenz, Kevin and Marleau, Peter and Mrowka, Stanley},
  journal={IEEE Transactions on Nuclear Science}, 
  title={Results With the Neutron Scatter Camera}, 
  year={2009},
  volume={56},
  number={3},
  pages={1269-1273},
  doi={10.1109/TNS.2009.2016659}}

@ARTICLE{nc3,
  title    = "Image reconstruction of a neutron scatter camera",
  author   = "Zhang, XianPeng and Zhang, Mei and Sheng, Lang and Zhang,
              ZhongBing and Li, KuiNian and Peng, BoDong and Zhang, XiaoDong
              and Ouyang, XiaoPing and Liu, Jun and Liu, JinLiang and Chen,
              Liang and Zhu, Jie and He, ChaHui",
  journal  = "Science China Technological Sciences",
  volume   =  59,
  number   =  1,
  pages    = "149--155",
  month    =  jan,
  year     =  2016
}

@INPROCEEDINGS{nc4,
  author={Greenberg, Charles H. and Brennan, James and Mascarenhas, Nicholas and Marleau, Peter and Mrowka, Stan},
  booktitle={2009 IEEE Nuclear Science Symposium Conference Record (NSS/MIC)}, 
  title={Exploring neutron scatter camera performance using {MCNP-PoliMi}}, 
  year={2009},
  volume={},
  number={},
  pages={1274-1276},
  doi={10.1109/NSSMIC.2009.5402371}}

@ARTICLE{nc5,
  author={Brennan, James and Brubaker, Erik and Cooper, Robert and Gerling, Mark and Greenberg, Charles and Marleau, Peter and Mascarenhas, Nicholas and Mrowka, Stanley},
  journal={IEEE Transactions on Nuclear Science}, 
  title={Measurement of the Fast Neutron Energy Spectrum of an $^{241}${Am}-{Be} Source Using a Neutron Scatter Camera}, 
  year={2011},
  volume={58},
  number={5},
  pages={2426-2430},
  doi={10.1109/TNS.2011.2163192}}

@article{nc6,
    author = {Goldsmith, John E. M. and Gerling, Mark D. and Brennan, James S.},
    title = {A compact neutron scatter camera for field deployment},
    journal = {Review of Scientific Instruments},
    volume = {87},
    number = {8},
    pages = {083307},
    year = {2016},
    month = {08},
    issn = {0034-6748},
    doi = {10.1063/1.4961111},
    url = {https://doi.org/10.1063/1.4961111},
}

@INPROCEEDINGS{nc7,
  author={Mascarenhas, Nick and Brennan, Jim and Krenz, Kevin and Lund, Jim and Marleau, Peter and Rasmussen, Julia and Ryan, Jim and Macri, John},
  booktitle={2006 IEEE Nuclear Science Symposium Conference Record}, 
  title={Development of a Neutron Scatter Camera for Fission Neutrons}, 
  year={2006},
  volume={1},
  number={},
  pages={185-188},
  doi={10.1109/NSSMIC.2006.356135}}

@Article{compton,
AUTHOR = {Parajuli, Raj Kumar and Sakai, Makoto and Parajuli, Ramila and Tashiro, Mutsumi},
TITLE = {Development and Applications of {Compton} Camera---A Review},
JOURNAL = {Sensors},
VOLUME = {22},
YEAR = {2022},
NUMBER = {19},
ARTICLE-NUMBER = {7374},
URL = {https://www.mdpi.com/1424-8220/22/19/7374},
PubMedID = {36236474},
ISSN = {1424-8220},
DOI = {10.3390/s22197374}
}

@article{spallation,
   author = "Fomin, Nadia and Fry, Jason and Pattie, Robert W. and Greene, Geoffrey L.",
   title = "Fundamental Neutron Physics at Spallation Sources", 
   journal= "Annual Review of Nuclear and Particle Science",
   year = "2022",
   volume = "72",
   pages = "151--176",
   doi = "10.1146/annurev-nucl-121521-051029",
   url = "https://www.annualreviews.org/content/journals/10.1146/annurev-nucl-121521-051029",
   publisher = "Annual Reviews",
   issn = "1545-4134",
   type = "Journal Article",
  }

@article{sp1,
    author = "Collar, J. I. and others",
    title = "{Coherent elastic neutrino-nucleus scattering at the Japan Proton Accelerator Research Complex}",
    eprint = "2512.19788",
    archivePrefix = "arXiv",
    primaryClass = "hep-ph",
    doi = "10.1007/JHEP05(2026)106",
    journal = "JHEP",
    volume = "05",
    pages = "106",
    year = "2026"
}

@article{sp2,
    author = "Abele, H. and others",
    title = "{Particle Physics at the European Spallation Source}",
    eprint = "2211.10396",
    archivePrefix = "arXiv",
    primaryClass = "physics.ins-det",
    doi = "10.1016/j.physrep.2023.06.001",
    journal = "Phys. Rept.",
    volume = "1023",
    pages = "1--84",
    year = "2023"
}

@article{sp3,
    author = "Baxter, D. and others",
    title = "{Coherent Elastic Neutrino-Nucleus Scattering at the European Spallation Source}",
    eprint = "1911.00762",
    archivePrefix = "arXiv",
    primaryClass = "physics.ins-det",
    reportNumber = "IFIC/19-45, YITP-SB-19-37, FERMILAB-PUB-19-612-V",
    doi = "10.1007/JHEP02(2020)123",
    journal = "JHEP",
    volume = "02",
    pages = "123",
    year = "2020"
}

@phdthesis{leire,
 title={Towards high-pressure noble gaseous detectors for coherent elastic neutrino-nucleus scattering}, 
 author={Leire Larizgoitia},
 year={2026},
 eprint={2603.15866},
 school={EHU/UPV},
 archivePrefix={arXiv},
 primaryClass={physics.ins-det},
 url={https://arxiv.org/abs/2603.15866}, 
}

@article{science,
	author = {Akimov, D. and others},
    title={Observation of Coherent Elastic Neutrino-Nucleus Scattering},
	volume = {357},
	number = {6356},
	pages = {1123--1126},
	year = {2017},
	doi = {10.1126/science.aao0990},
	publisher = {American Association for the Advancement of Science},
	issn = {0036-8075},
	journal = {Science},
    eprint = "1708.01294",
    archivePrefix = "arXiv",
}

@phdthesis{bjorn,
    author = "Scholz, Bjorn Jorg",
    title = "{First Observation of Coherent Elastic Neutrino-Nucleus Scattering}",
    eprint = "1904.01155",
    archivePrefix = "arXiv",
    primaryClass = "nucl-ex",
    doi = "10.1007/978-3-319-99747-6",
    school = "Chicago U.",
    year = "2017"
}

@article{2761,
  title   = {Characterization of {EJ-276D} plastic scintillator and its comparison with {EJ-299-33A} and {BC-501A}},
  author  = {Pant, Pankaj and Banerjee, K. and Roy, P. and Shil, R. and Saha, A. K.},
  journal = {JINST},
  volume  = {19},
  number  = {10},
  pages   = {P10036},
  year    = {2024},
  doi     = {10.1088/1748-0221/19/10/P10036}
}

@article{2762,
title = {Discrimination of neutron-gamma in the low energy regime using machine learning for an {EJ-276D} plastic scintillator},
journal = {Nucl. Instrum. Meth. A},
volume = {1083},
pages = {171170},
year = {2026},
issn = {0168-9002},
doi = {10.1016/j.nima.2025.171170},
url = {https://www.sciencedirect.com/science/article/pii/S0168900225009726},
author = {S. Panda and P.K. Netrakanti and S.P. Behera and R.R. Sahu and K. Kumar and R. Sehgal and D.K. Mishra and V. Jha}
}

@article{psd1,
title = {An artificial neural network based neutron--gamma discrimination and pile-up rejection framework for the {BC}-501 liquid scintillation detector},
journal = {Nucl. Instrum. Meth. A},
volume = {610},
number = {2},
pages = {534-539},
year = {2009},
issn = {0168-9002},
doi = {10.1016/j.nima.2009.08.064},
url = {https://www.sciencedirect.com/science/article/pii/S0168900209017045},
author = {E. Ronchi and P.-A. Soderstrom and J. Nyberg and E. {Andersson Sunden} and S. Conroy and G. Ericsson and C. Hellesen and M. {Gatu Johnson} and M. Weiszflog}
}

@article{psd2,
title = {Test of digital neutron--gamma discrimination with four different photomultiplier tubes for the {NEutron Detector Array (NEDA)}},
journal = {Nucl. Instrum. Meth. A},
volume = {767},
pages = {83-91},
year = {2014},
issn = {0168-9002},
doi = {10.1016/j.nima.2014.08.023},
url = {https://www.sciencedirect.com/science/article/pii/S0168900214009437},
author = {X.L. Luo and V. Modamio and J. Nyberg and J.J. Valiente-Dobon and Q. Nishada and G. {de Angelis} and J. Agramunt and F.J. Egea and M.N. Erduran and S. Erturk and G. {de France} and A. Gadea and V. Gonzalez and T. Huyuk and G. Jaworski and M. Moszynski and A. {Di Nitto} and M. Palacz and P.-A. Soderstrom and E. Sanchis and A. Triossi and R. Wadsworth}
}

@phdthesis{drew,
author={Fustin, Drew A.},
title={First Dark Matter Limits from the {COUPP} 4kg Bubble Chamber at a Deep Underground Site},
school={University of Chicago},
year={2012},
eprint         = "2401.07384",
archivePrefix  = "arXiv",     
}

@article{ficenec,
    author = "Awe, C. and Barbeau, P. S. and Collar, J. I. and Hedges, S. and Li, L.",
    title = "{Liquid scintillator response to proton recoils in the 10{\textendash}100 keV range}",
    eprint = "1804.06457",
    archivePrefix = "arXiv",
    primaryClass = "physics.ins-det",
    doi = "10.1103/PhysRevC.98.045802",
    journal = "Phys. Rev. C",
    volume = "98",
    number = "4",
    pages = "045802",
    year = "2018"
}

@article{HU201978,
title = {Measurements of cosmic ray induced background neutrons near the ground using a {Bonner} sphere spectrometer},
journal = {Nucl. Instrum. Meth. A},
volume = {940},
pages = {78-82},
year = {2019},
issn = {0168-9002},
doi = {10.1016/j.nima.2019.06.004},
url = {https://www.sciencedirect.com/science/article/pii/S0168900219308216},
author = {Z.M. Hu and L.J. Ge and J.Q. Sun and Y.M. Zhang and Z.Q. Cui and G. Gorini and H. Zhang and J. Chen and J.X. Chen and X.Q. Li and T.S. Fan}
}

\end{document}